\documentclass[submission,copyright,creativecommons]{eptcs}
\providecommand{\event}{SynCoP 2014} 

\usepackage{subfig}
\usepackage[utf8]{inputenc}
\usepackage[noadjust]{cite}
\usepackage{url}
\usepackage{graphicx}

\usepackage{amsmath}
\usepackage{amsthm}
\usepackage{amssymb}
\usepackage{tikz}
\usetikzlibrary{automata}

\tikzset{
    state/.style={
           rectangle,
           rounded corners,
           draw=black, 
           fill=green!5,
           minimum height=2em,
           inner sep=2pt,
           text centered,
           font=\normalfont,
           },
}
\tikzset{edge/.style={left, font={\sffamily\small}}}
\tikzstyle{line} = [draw, -latex']

\newtheorem{definition}{Definition}

\newcommand{\A}{\mathcal{A}}
\newcommand{\IM}{\ensuremath{\mathrm{IM}}}

\newcommand{\Param}{U} 

\newcommand{\py}{\pi} 

\newcommand{\slope}{\mathit{slope}}

\newcommand{\steps}[0]{ {\rightarrow} }

\newcommand{\colorevent}{blue}

\graphicspath{{figures/}}

\ifdefined \VersionWithComments
	\usepackage{marginnote}
	
\newcommand{\marginX}{\marginnote{\huge{\quad\quad\textbf{!}\quad\quad}}}
	\newcommand{\commentaire}[1]{{\color{red}\marginX{}$\Leftarrow$ 
\textbf{#1} $\Rightarrow$}}
	\newcommand{\ea}[1]{{\color{violet}\marginX{}[\textbf{\'Etienne}: #1]}}
	\newcommand{\gl}[1]{{\color{magenta}\marginX{}[\textbf{Giuseppe}: #1]}}
	\newcommand{\yc}[1]{{\color{green}\marginX{}[\textbf{Youcheng}: #1]}}
\else
	\newcommand{\commentaire}[1]{}
	\newcommand{\ea}[1]{}
	\newcommand{\gl}[1]{}
	\newcommand{\yc}[1]{}
\fi

\title{Toward Parametric Timed Interfaces for Real-Time Components}
\def\titlerunning{Toward Parametric Timed Interfaces for Real-Time Components}

\author{Youcheng Sun$^{1,2}$, Giuseppe Lipari$^{1,2}$, \'Etienne André$^{3}$ and Laurent Fribourg$^{2}$
\institute{$^1$Scuola Superiore Sant'Anna, Pisa, Italy\thanks{%
	The research leading to these results has received funding from the European Union Seventh Framework Programme (FP7/2007-2013) under grant agreement No.\ 246556.
}\\
$^2$LSV, ENS Cachan \& CNRS, France\\
$^3$Université Paris 13, Sorbonne Paris Cité, LIPN, CNRS, UMR 7030, F-93430, Villetaneuse, France\thanks{%
	This work is partially supported by STIC Asie project CATS (``Compositional Analysis of Timed Systems'').
}}}

\def\authorrunning{Y. Sun, G. Lipari, \'E. André \& L. Fribourg}

\begin{document}

\maketitle

\ifdefined \VersionWithComments
	\textcolor{red}{\textbf{This is the version with comments. To disable comments, comment out line~3 in the \LaTeX{} source.}}
\fi


\begin{abstract}
We propose here a framework to model real-time components consisting of concurrent real-time tasks running on a single processor, using parametric timed automata. Our framework is generic and modular, so as to be easily adapted to different schedulers and more complex task models.
We first perform a parametric schedulability analysis of the components using the inverse method. We show that the method unfortunately does not provide satisfactory results when the task periods are considered as parameters. After identifying and explaining the problem, we present a solution adapting the model by making use of the worst-case scenario in schedulability analysis. We show that the analysis with the inverse method always converges on the modified model when the system load is strictly less than~100\%. 
Finally, we show how to use our parametric analysis for the generation of timed interfaces in compositional system design.
\end{abstract}
%
%


\section{Introduction}
\label{sec:intro}


\yc{Content in abstract is repeated here ...}
Designing and analysing distributed real-time systems is a very
challenging task. The main source of complexity arises from the large
number of parameters to consider: tasks priorities, computation times
and deadlines, synchronisation, precedence and communication
constraints, etc. Finding the optimal values for the parameters is not
easy and often a small change in one parameter may completely change
the behaviour of the system and even compromise its correctness. For
these reasons, designers are looking for analysis methodologies that
allow incremental design and exploration of the parameter space.

We consider here real-time systems consisting of a set of real-time 
tasks executed concurrently on a single processor platform.
Each task can be time-triggered or event-triggered: in the first case, 
it is activated periodically, and each time it executes a portion of code 
called \emph{job} or \emph{instance}, after which it self-suspends, 
waiting for their next periodic activation. 
In the second case, instances are
activated by internal or external events. Each task is characterised 
by a \emph{relative deadline}, that is the maximum amount of time 
that must elapse from the activation of one instance to its 
completion.

%
A \emph{scheduler} is needed to decide which 
task to execute at each instant. The scheduler can be \emph{on-line} 
if the decision is taken while the system is running depending on 
the current state; or \emph{off-line} if the schedule is pre-computed
before the system runs. 
Fixed Priority Preemptive Scheduling (FPPS) has been standardised in 
POSIX \cite{POSIX-RT} and is currently available in all commercial 
and open-source Real-Time Operating Systems.

One important requirement of real-time systems is to ensure that 
the system is \emph{schedulable}, i.e.\ that all tasks will always 
complete before their deadlines when scheduled by the selected algorithm.
Testing the system under different input and state conditions does not 
guarantee the system \emph{schedulability} (i.e.\ that the system is 
schedulable), because the number of possibilities to test is too large to 
guarantee complete coverage of all possible cases. A better approach is to 
build an abstract model of the system, and perform analysis on the model.

A large body of research literature has addressed the problem of
schedulability analysis of real-time tasks, both using formal methods 
(e.g.\ \cite{AM02,eta:edf,eta:2clocks}) and mathematical equations
(e.g.\ \cite{Buttazzo-Book-2004,liu2000real}).
In the literature, a task is typically modelled by several parameters, 
typically 
(i) a \emph{worst-case computation time} (i.e.\ an upper bound of 
the execution time of every instance of the task under every possible 
condition), 
(ii) a \emph{deadline}, and (iii) an \emph{activation pattern} (e.g.\ 
periodic, sporadic, arrival curve). 
A periodic task is activated every period; a sporadic task can be activated 
at any time, but the distance between two activations is lower bounded by a 
constant \emph{minimum interarrival time};
finally, an arrival curve \cite{thiele2000real} is a function $\alpha(t)$ that defines an upper bound on the maximum number of activations in any interval of length~$t$.

\subsubsection*{Real-Time Components and Timed Interfaces}

For complex distributed real-time systems, a component-based methodology
may help reduce the complexity of the design and analysis phases. This 
paper is a first step toward the definition of a \emph{timed 
interface} for a real-time component. Therefore, we now describe our 
notion of real-time component and timed interface. 

We define a \emph{distributed real-time system} as a set of \emph{real-time components}. Each component runs on a dedicated single processor node, and all components are connected to each other by a local network. 
A component consists of a \emph{provided interface}, a \emph{required 
interface}, and an \emph{implementation} (see e.g. \cite{Rahmani:2011:SMD:1988997.1989008}).

The \emph{provided interface} is a set of methods that a component makes 
available to other components of the system. Each method is 
characterised by: (i) the method signature, which is the name of
the method and the list of parameters, and (ii) a worst-case
activation pattern, which describes the maximum number
of invocations the method is able to handle in any interval
of time. In this paper, we will describe the worst-case activation 
pattern by an \emph{arrival curve}~\cite{thiele2000real}.
\ea{strange reference, I could not find it on DBLP! What is the first name of the authors?}\gl{The first author is Lothar Thiele. The bibtex file was corrupted, I corrected it. It is a famous paper with 376 citations according to google scholar, but maybe it is not referenced on DBLP.}
The semantic of invocation of a method can be synchronous (the caller waits 
for the method to be completed) or asynchronous (the caller continues to 
execute without waiting for the completion of the operation).

The \emph{required interface} is a set of methods that the component 
requires for carrying out its services. Each method is characterised by its 
signature and a worst-case invocation pattern.

The \emph{implementation} of a component is the specification
of how the component carries out its work. In our model,
a component is implemented by a set of concurrent real-time tasks and by 
a scheduler. Tasks can be time-triggered, when periodically activated; or 
event-triggered, in which case they are activated by a call to a provided
method of the component. In other words, an event-triggered task 
implements one method of the provided interface, and in turn it may 
invoke a method of the required interface. 

\begin{figure}
\begin{center}
\includegraphics[width=6cm]{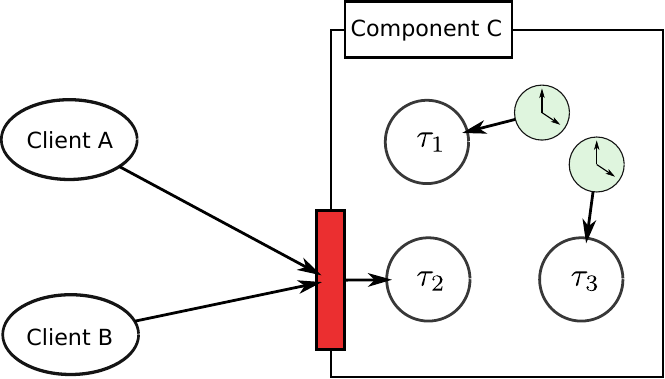}
\end{center}
\caption{A component with three tasks and one method in the provided interface.}
\label{fig:component-representation}
\end{figure}

A graphical representation of a component is shown in 
Figure~\ref{fig:component-representation}. 
\gl{It may be the case to add a figure with many components connected to each other via interfaces.}%
In this example, the component provides one single method in the 
provided interface (pictorially represented by the red rectangle), and 
does not specify any method in the required interface. The component is 
implemented by three tasks: tasks $\tau_1$ and $\tau_3$ are time 
triggered (the green clocks in the picture), whereas task $\tau_2$ implements the method in the provided interface, and hence it is triggered by invocations from external clients. 

In a \emph{component-based} design methodology, components are 
independently designed and developed, and then \emph{integrated} in the 
final system by connecting them together through their interfaces. 
It is clear that the interface specification plays an important role in 
this methodology: for a real-time component, the interface should contain 
not only the functional specification (i.e.\ method signature, constraints 
on the parameters, etc.) but also the \emph{timed behaviour} of the 
component. In particular, in this paper we enhance the specification of the 
interface by adding parameters on the activation pattern and on the 
response delay of a method. 

Given a component $\mathcal{C}$, its \emph{provided interface} is thus 
defined as:
\begin{itemize}
\item a set of method signatures $m_1, m_2, \ldots$;
\item a \emph{parametric arrival curve} $\alpha_i(t)$ for each method 
$m_i$, which represents the activation pattern that the corresponding 
implementing task will receive;
\item a \emph{worst-case response time} $D_i$ parameter for each method 
$m_i$.
\end{itemize}

Similarly, the \emph{required interface} of a component is defined as:
\begin{itemize}
\item a set of method signatures $m_1, m_2, \ldots$;
\item a \emph{parametric arrival curve} $\alpha_i(t)$ for each method 
$m_i$ that represents the activation pattern generated by this component;
\item for every synchronous method call, a maximum allowed delay $R_i$ in 
receiving the response.
\end{itemize}

Finally, the component is characterised by a \emph{set of constraints} on 
the parameters: for all valuations of the parameters satisfying the 
constraints, the component is guaranteed to be correct both from the 
functional point of view (i.e.\ the component produces correct values) and 
from the timing point of view (i.e.\ all tasks complete before their 
deadlines, and all provided functions return their values within the 
desired maximum response delay).  

The road to realise such a component-based design methodology is long and 
many theoretical and practical problems need to be solved before the 
methodology can be used in practice. One important problem is how to 
compute the set of constraints that define the correct behaviour of a 
component. In the process of designing and analysing a component in 
isolation, it is necessary to use parametric arrival curves for describing 
the activation patterns for event-triggered tasks, and parametric deadlines 
for bounding their response times. Performing a parametric analysis aims at 
deriving a set of constraints for these parameters that make the component 
schedulable. During integration, the correctness of the system is checked by 
intersecting the constraints of the communicating components to see if there 
is some feasible assignment of parameters that makes all components 
schedulable.

\subsubsection*{Objectives}

Our general research agenda, that goes beyond the scope of this paper,
is to establish a component-based design methodology and analysis for real-
time components. One of the important steps in the proposed methodology is 
to be able to perform parametric analysis of a component with respect to its 
activation patterns.

In this paper, we focus on solving two specific problems: 
\begin{enumerate}
	\item how to build a formal parametric model of a component consisting 
of a set of real-time tasks, some of which can be periodic, others can be 
activated by generic arrival curves; and
	\item how to perform a parametric analysis of the schedulability of the 
component, thus deriving a set of constraints that define the space of 
parameters that make the component schedulable. 
\end{enumerate}

For the sake of simplicity, in this paper we focus only on the
\emph{provided interface} of a component; that is, we investigate on the 
parametric analysis of a component with respect to the patterns of 
activations. The analysis of the required interface is the subject of future work.

\subsubsection*{Contributions}

%

Our contribution is threefold.
First, we propose a formal model of a real-time component based
on parametric timed automata~\cite{AHV93}, a popular formalism for modelling real-time systems. 
Unlike many similar models proposed in the literature, our modelling framework 
is completely modular: a system is obtained by combining simpler 
automata, each one implementing one aspect of the component. In 
particular, we separate the specification of the task behaviour
from the activation pattern (periodic, sporadic or generic arrival 
curve), and from the scheduler. In this way we can easily and seamlessly 
change the scheduler and the activation pattern of a task without 
changing the rest of the component specification, which is very important during the parametric analysis of a component. 
For the analysis, we use the inverse method~\cite{AS13} and the IMITATOR tool~\cite{AFKS12}.

As a second contribution, we show that, when the activation patterns are parametric, the inverse method does not provide satisfactory results, in the sense that it may output a constraint reduced to a single point.
We describe the problem and provide a solution that is valid for periodic 
(with no offset), sporadic and generic activation patterns that can be 
described by arrival curves. 

Finally, as third contribution, 
we describe how this model can be used as a basis for synthesising the timed interface of a real-time component. 

\subsubsection*{Organisation of the Paper}

The rest of this paper is organised as follows.
Section~\ref{sec:related} reviews related work.
Section~\ref{sec:preliminaries} recalls the necessary preliminaries, viz.\ real-time systems, parametric timed automata and the inverse method.
Section~\ref{sec:model-rts} presents our model of a real-time component using parametric timed automata.
Section~\ref{sec:parametric-analysis} introduces our parametric analysis, allowing to deal with parametric 
task activations\ea{?}.
Section~\ref{sec:interface} introduces preliminary work allowing to perform a component-based parametric analysis.
Section~\ref{sec:conclusions} concludes and present further directions of research.

\section{Related Work}
\label{sec:related}

Parametric analysis of real-time systems using mathematical equations 
has already been addressed in the past. Bini et al.~\cite{Bin06} 
proposed a method for parametric analysis of real-time periodic tasks 
where parameters can be either worst-case computation times or task 
periods.  However, with Bini et al.'s approach, changing the task model 
requires the development of a new methodology.

A more general approach to scheduling analysis is to use formal methods for modelling a real-time system. A formal framework for scheduling problems using timed automata with stopwatches has been proposed in~\cite{AM02}.
Fersman et al.\ \cite{eta:2clocks} proposed a \emph{Task Automaton}. Similar approaches have been proposed using time(d) Petri nets \cite{bucci2004timed,Lime2009}.

It is possible to perform an exploration of the parameter space using 
timed automata, as in \cite{Pal13}. However, their approach is not fully 
parametric: the analysis is repeated for all combination of the discrete 
values of the parameters. Hence, their method does not scale well as the 
number of parameters and the number of discrete values increases.
Furthermore, that approach does not consider non-integer points, and 
cannot be used to quantify the system robustness.

Full parametric analysis can be performed using specific formalisms. 
For example, formalisms such as parametric timed automata (PTA)~\cite{AHV93}
and parametric time Petri nets~\cite{TLR09}, have been used to model 
parametric schedulability problem (see, e.g.~\cite{CPR08,SSLAF13}).
In particular, thanks to generality of these modelling languages, it is 
possible to model a larger class of constraints, and perform full parametric  
analysis on many different variables, for example task offsets. 

The inverse method~\IM{}~\cite{AS13} can be used for exploring the space of 
parameters of a parametric timed automaton (and, more generally, of a parametric \emph{stopwatch} automaton) in the proximity of a valuation point.
In this paper we use this method for performing parametric analysis of real-time systems where task activation patterns are modelled with parametric \emph{arrival curves}. 

We have used a similar approach in \cite{SSLAF13}, where a distributed 
real-time system has been modelled using parametric stopwatch automata.
However, in \cite{SSLAF13}  the methodology is limited to only use the tasks' computation times as  parameters.
Here, we investigate a situation where arrival curves are considered as parameters too.
Furthermore, our final goal is to be able to perform interface-based parametric analysis.

Our generic modular approach can be seen as a contract-based methodology where ``provided'' and ``required'' interfaces are instances of (assumption, guarantee) pairs in the contract terminology. An interface-based approach to the design and analysis of real-time systems using assume/guarantees has already been proposed in the literature \cite{shin2008compositional,lampka2013component}, but their approach is not parametric. 
Compositional verification of timed systems, using assume guarantee reasoning, has also been considered in~\cite{LALSD14} for event-recording automata, a subclass of timed automata; again, this approach is non-parametric.

\section{Preliminaries}
\label{sec:preliminaries}

\subsection{Real-Time Tasks}
\label{sec:task-model}

A real-time task $\tau_i$ is a sequence of instances (or jobs) 
$J_{i,k}$, with $k=0, 1, \ldots$.
Each instance $J_{i,k} = (a_{i,k}, 
c_{i,k}, d_{i,k})$ is characterised by an arrival time $a_{i,k}$, a 
computation time $c_{i,k}$, and an absolute deadline $d_{i,k}$.
The system is schedulable if the scheduling algorithm orders the 
execution times of the jobs such that each job executes $c_{i,k}$ 
units of execution in interval $[a_{i,k}, d_{i,k}]$. Additionally, an 
instance can only start executing after the previous instances from the same 
task have completed: if we denote by $f_{i,k-1}$ the finishing time of the 
$(k-1)$th instance, then each job can only execute in interval 
$[\max(f_{i,k-1}, a_{i,k}), d_{i,k}]$. 

Task $\tau_i$ is then characterised by three parameters:
\begin{itemize}
\item the Worst-Case Execution Time $C_i$, which is an upper bound on 
the execution time of any instance of the task
(i.e.\ $\forall k> 0: c_{i,k} \leq C_i $);
\item the \emph{relative deadline} $D_i$; the absolute deadline of 
every instance can be computed as $d_{i,k} = a_{i,k} + D_i$;
\item the arrival pattern.
\end{itemize}

For the arrival pattern, we consider three kinds of schemes: 
\begin{itemize}
	\item \textbf{Periodic}: this arrival pattern is characterised by a \emph{period} 
	$T_i$, and the arrival time of every instance is computed as: 
	\begin{align*}
	a_{i,0} &= 0 \\
	\forall k > 0: a_{i,k} &= a_{i,k-1} + T_i 
	\end{align*}
	
	\item \textbf{Sporadic}: this arrival pattern is characterised by a \emph{minimum 
	interarrival time} that we denote again by~$T_i$, and the arrival 
	times of every instance must respect the following constraints:
	\begin{align*}
	a_{i,0} &= 0 \\
	\forall k > 0 : a_{i,k} &\geq a_{i,k-1} + T_i 
	\end{align*}
	
	\item \textbf{Arrival curve} \cite{thiele2000real}: 
	in this case the 
	pattern of arrival must respect a certain function called 
	\emph{arrival curve} $\alpha_i(t) : \mathbb{R} \rightarrow 
	\mathbb{N}$.
	The arrival curve constrains the number~$n$ of arrivals in 
	any interval of a given length~$\Delta$:
	\[
	\forall k \geq 0, \forall n > 0 : n \leq \alpha_i(a_{i,k+n-1} - a_{i,k}) 
	\]
	In other words, the number of arrival events in any interval must not 
	exceed the value of the arrival curve for that interval\footnote{Unlike 
	in \cite{thiele2000real}, for simplicity in this paper we only consider 
	upper bound arrival curves.}.
	Arrival curves are monotonically non-decreasing, and convex, 
	i.e.\ $\forall t, \delta  :\alpha_i(t + \delta) \leq \alpha_i(t) 
	+ \alpha_i(\delta)$.
	The value of the an arrival curve at time $0$
	is also called \emph{burstiness} and represents the amount of 
	simultaneous arrival 
	events that can be sent to a task.  
	Arrival curves are a generalisation of the sporadic arrival model.
	In  fact, a sporadic task can be represented by an arrival curve with 
	burstiness $\alpha_i(0) = 1$ and a periodic behaviour. However, an 
	arrival curve can have any convex shape.

	The sum of two arrival curves is still an arrival curve. Also, we can 
	define a partial order relationship between arrival curves using the 
	natural ordering between values of the function: 
	$\alpha_i(\cdot) \preceq \alpha_j(\cdot)$ iff $\forall t \; \alpha_i(t) \preceq \alpha(t)$. 

	In this paper we deal with parametric arrival curves. In 
	particular, we will use periodic arrival curves of the form: 
	\begin{equation}
	\label{eq:param-arr-curve}
		\alpha_{N^u, P}(t) = N^u + \left\lfloor \frac{t}{P} \right\rfloor
	\end{equation}
	where $N^u$ is a discrete parameter that denotes the initial 
	burstiness, and $P$ is a continuous parameter that denotes the 
	period. Using the partial order relationship, a generic arrival 
	curve can always be upper bounded by a periodic arrival curve of 
	the form (\ref{eq:param-arr-curve}).

\end{itemize}

\subsection{Parametric Stopwatch Automata}
\label{sec:pta}

We introduce here an extension of parametric timed automata that will be 
used in Section~\ref{sec:model-rts} to model real-time systems.
Timed automata are finite-state automata augmented with clocks, i.e.\ 
real-valued variables increasing uniformly, that are compared within
guards and invariants with timing delays~\cite{AD94}. Parametric 
timed automata (PTA)~\cite{AHV93} extend timed automata with
parameters, i.e.\ unknown constants, that can be used in guards and 
invariants. We will use here an extension of PTA with 
\emph{stopwatches}~\cite{AM02}, where clocks can be stopped in some 
control states of the automaton.

	\ea{I have the impression that this section goes too much into details, since most of the notations are not used afterwards; anyway, due to the short time, I leave it as it is.}
Given a set~$X$ of clocks and a set~$\Param$ of parameters, a
constraint~$C$ over~$X$ and~$\Param$ is a conjunction of linear
inequalities on~$X$ and~$\Param$\footnote{%
	Note that this is a more general form than the strict original 
definition of PTA~\cite{AHV93};
	since most problems for PTA are undecidable anyway, this has no 
practical incidence, and increases the expressiveness of the formalism.
}.  Given a parameter valuation (or
point)~$\py$, we write~$\py \models C$ when the constraint where all
parameters within~$C$ have been replaced by their value as in~$\py$ 
is satisfied by a non-empty set of clock valuations.

\begin{definition}
A parametric timed automaton with stopwatches (PSA)~$\A$ is \mbox{$(\Sigma, Q, 
q_{0}, X, \Param, K, I, \slope, \steps)$} with
	$\Sigma$ a finite set of actions,
	$Q$ a finite set of locations,
	$q_{0} \in Q$ the initial location,
	$X$ a set of $h$ clocks,
	$\Param$ a set of parameters,
	$K$ a constraint over~$\Param$,
	$I$ the invariant assigning to every $q\in Q$ a constraint over~$X$ 
and~$\Param$,
	$\mathit{slope}:Q \rightarrow \{ 0, 1 \}^h$ assigns a constant slope to 
every location,
	and
	$\steps$ a step relation consisting of elements $(q,g,a,\rho,q')$, 
	where
	$q,q'\in Q$, $a\in\Sigma$, $\rho\subseteq X$ is the set of clocks to be 
reset, and
	the guard $g$ is a constraint over~$X$ and~$\Param$.
\end{definition}

The $\mathit{slope}$ function is the extension of parametric timed automata to \emph{stopwatch} timed automata, since it allows one to stop the time elapsing of some clock variables in some locations.
This expressive power is used in the context of schedulability to model the preemption mechanism.

It is well-known that the parallel composition (using a synchronisation on actions) of several PSA is itself a PSA.
Hence, it is common to model a complex system by composing several system components modelled themselves using PSA.
 
The semantics of a PSA~$\A$ is defined in terms of states, i.e.\ 
pairs $(q, C)$ where~$q \in Q$ and~$C$ is a constraint over~$X$
and~$\Param$.
Given a point~$\py$, we say that a state $(q, C)$ is $\py$-compatible 
if $\py \models C$.
Runs are alternating sequences of
states and actions, and traces are time-abstract runs, i.e.\ 
alternating sequences of \emph{locations} and actions. The trace set
of~$\A$ corresponds to the traces associated with all the runs
of~$\A$. Given~$\A$ and~$\py$, we denote by~$\A[\py]$ the
(non-parametric) timed stopwatch automaton where each occurrence of a parameter has been replaced by its constant value as in~$\py$.
%
%
Details can be found in, e.g.~\cite{AS13}.

\subsection{The Inverse Method}
\label{sec:im}

The inverse method for PSA~\cite{AS13} exploits the
knowledge of a reference point of timing values for which the good
behaviour of the system is known. The method synthesises automatically
a dense space of points around the reference point, for which the
discrete behaviour of the system, that is the set of all the
admissible sequences of interleaving events, is guaranteed to be the
same.

The inverse method~$\IM$ proceeds by exploring iteratively longer runs 
from the initial state.
When a $\py$-incompatible state is met (that is a state $(q, C)$ such that 
$\py \not\models C$),
a $\py$-incompatible inequality~$J$ is selected within the projection of~$C$ 
onto~$\Param$.
This inequality is then negated, and the analysis restarts with a model further 
constrained by~$\neg J$.
When a fixpoint is reached, that is when no $\py$-incompatible state is found 
and all states have their successors within the set of reachable states, the 
intersection of all the constraints onto the parameters is returned.

$\IM$ proceeds by iterative state space
exploration, and its result comes under the form of a fully
parametric constraint.
By repeatedly applying the method, we are able to decompose the 
parameter space into a covering set of ``tiles'', which ensure a 
uniform behaviour of the system: it is sufficient to test only one 
point of the tile in order to know whether or not the system behaves 
correctly on the whole tile. This is known as the \emph{behavioural 
cartography}~\cite{AF10}.
Both the inverse method and the behavioural cartography are semi-algorithms;
that is, they are not guaranteed to terminate but, if they do, their result is correct.

\section{A Modular Framework for Modelling Real-Time Systems}
\label{sec:model-rts}


In this section we refer to a real-time system as a set of 
real-time tasks scheduled by a FPPS on a single processor.
Of course, the discussion is valid also when considering a single component of a large real-time distributed system. 

Our model of a real-time system consists of three kinds of PSA components: 
the task automata, the task activation automata and the scheduler
automaton.
We refer to the composition of these PSA components through synchronisation labels as the \textsf{system automaton}.
\ea{I think there are more than three components; it should rather be three kinds of PSA components (because there is more than one task!)}
\gl{ok}

\begin{figure}[t]
        \captionsetup{justification=centering}
	\centering

	\subfloat[Task automaton]{
		\scalebox{0.75}{
			  \begin{tikzpicture}[->, =stealth!]

    \node[initial above, initial text={}, state, text width=2.5cm, 
anchor=center] at (8, 8) (IDLE) {      
      \begin{tabular}{c}
        \textbf{Idle} \\
      \end{tabular}
         
    };

    \node[state, accepting, text width=2.5cm, anchor=center] at (0, 8) 
(ACTEVENT) {
      \begin{tabular}{c}
        \textbf{ActEvent} \\
      \end{tabular}
         
    };

    \node[state, text width=2.5cm] at (0, 4) (WAITING) {
      \begin{tabular}{c}
        \textbf{Waiting}\\ 
        $\mathsf{stop}\{c\}$\\
        $d \leq D$
      \end{tabular}
    };

    \node[state, text width=2.5cm] at (8, 4) (EXEC) {
      \begin{tabular}{c}
        \textbf{Running}\\ 
        $c \leq N*C \wedge $\\
        $d \leq D$
      \end{tabular}
    };

    \node[state, text width=2.5cm] at (4, 0) (DMISS) {
      \begin{tabular}{c}
        \textbf{Deadline}\\ 
        \textbf{Missed}
      \end{tabular}
    };

    \path 
    (IDLE)
    edge[] node[below] {
      \begin{tabular}{c}
        {\color{\colorevent} arrival\_event} \\
        $d := 0, c:= 0, N := 1$ 
      \end{tabular}
    } 
    (ACTEVENT)
    ;
 
    \path 
    (ACTEVENT)
    edge[] node[left] {
      \begin{tabular}{c}
        {\color{\colorevent} arrival} 
      \end{tabular}
    } 
    (WAITING)
    ;
 
    \path
    (WAITING)
    edge [below, out=-15, in=195] node {
      {\color{\colorevent} dispatch}
    }
    (EXEC)

    (WAITING)
    edge [right] node[right] {
      $ d = D $
    }
    (DMISS)
    ;

    \path
    (EXEC)
    edge [above, out=165, in=15] node {
      \begin{tabular}{@{} c @{}}
         {\color{\colorevent} preemption}\\
         $c < N*C$
      \end{tabular}
    }
    (WAITING)

    (EXEC)
    edge [left] node[left]{
      \begin{tabular}{c}
         $ c < N*C \wedge$ \\
         $d = D$
      \end{tabular}
    }
    (DMISS)

    (EXEC)
    edge [right] node[] {
      \begin{tabular}{@{} c @{}}
         {\color{\colorevent} end}\\
         $ c = N*C $
      \end{tabular}
    }
    (IDLE)
    ;

    \path
    (WAITING)
    edge [below, out=-90, in=-105, looseness=8] node[below left=-4mm] {
      \begin{tabular}{@{} c @{}}
         {\color{\colorevent} arrival\_event}\\
         $ d < D $ \\
         $ d := 0, N := N + 1$ 
      \end{tabular}
    }
    (WAITING)
    ;
    \path
    (WAITING)
    edge [below, out=-200, in=-215, looseness=8] node[above] {
      \begin{tabular}{@{} c @{}}
         {\color{\colorevent} arrival\_event}\\
      \end{tabular}
    }
    (WAITING)
    ;

    \path
    (EXEC)
    edge [below, out=-90, in=-75, looseness=8] node[below right=-4mm] {
      \begin{tabular}{@{} c @{}}
         {\color{\colorevent} arrival\_event}\\
         $ d < D $ \\
         $ d := 0, N := N + 1$ 
      \end{tabular}
    }
    (EXEC)
    ;
    \path
    (EXEC)
    edge [below, out=20, in=35, looseness=8] node[above] {
      \begin{tabular}{@{} c @{}}
         {\color{\colorevent} arrival\_event}\\
      \end{tabular}
    }
    (EXEC)
    ;

  \end{tikzpicture} 
		}
		\label{fig:task}
	}
	
	\subfloat[Task activation automaton]{
		\scalebox{0.75}{
						\begin{tikzpicture}[->, =stealth!]

				\node[initial left, initial text={}, state, text width=2.5cm, 
			anchor=center] 
			(ARREVENT) {      
				\begin{tabular}{c}
					\textbf{ArrEvent} \\
					$p\leq T$
				\end{tabular}
					
				};

				\path
				(ARREVENT)
				edge [above, out=105, in=75, looseness=12] node[right] {
				\begin{tabular}{@{} c @{}}
					{\color{\colorevent} arrival\_event}\\
					$ p = T $ \\
					$ p := 0 $ 
				\end{tabular}
				}
				(ARREVENT)
				;

			\end{tikzpicture} 
		}
	\label{fig:act}
	}
	\subfloat[Scheduler automaton]{
		\scalebox{0.75}{
		                     
  \begin{tikzpicture}[->, =stealth!]

    \node[initial right, initial text={}, state, anchor=center] at (8, 8) 
(IDLE) 
{      
      \begin{tabular}{c}
        \textbf{Idle} \\
      \end{tabular}
         
    };

    \node[state, accepting, anchor=center] at (6, 6) (ACT1) {
      \begin{tabular}{c}
        \textbf{Act1} \\
      \end{tabular}
         
    };

    \node[state, accepting, anchor=center] at (6, 10) (ACT2) {
      \begin{tabular}{c}
        \textbf{Act2} \\
      \end{tabular}
         
    };

    \node[state, anchor=center] at (3, 6) (RT1) {
      \begin{tabular}{c}
        \textbf{Rt1} \\
      \end{tabular}
         
    };

    \node[state, anchor=center] at (3, 10) (RT2) {
      \begin{tabular}{c}
        \textbf{Rt2} \\
      \end{tabular}
         
    };

    \node[state, accepting, anchor=center] at (3, 8) (ET1WT2) {
      \begin{tabular}{c}
        \textbf{Et1Wt2} \\
      \end{tabular}
         
    };

    \node[state, accepting, anchor=center] at (6, 8) (ET) {
      \begin{tabular}{c}
        \textbf{Et} \\
      \end{tabular}
         
    };
    %
    %
    \node[state, accepting, anchor=center] at (0, 10) (AT1RT2) {
      \begin{tabular}{c}
        \textbf{At1Rt2} \\
      \end{tabular}
         
    };

    \node[state, anchor=center] at (0, 6) (RT1WT2) {
      \begin{tabular}{c}
        \textbf{Rt1Wt2} \\
      \end{tabular}
         
    };

    \path 
    (IDLE)
    edge[] node[below right] {
      \begin{tabular}{c}
        {\color{\colorevent} $\mathsf{arrival_1}$} \\
      \end{tabular}
    } 
    (ACT1)
    ;

    \path 
    (IDLE)
    edge[] node[above right] {
      \begin{tabular}{c}
        {\color{\colorevent} $\mathsf{arrival_2}$} \\
      \end{tabular}
    } 
    (ACT2)
    ;

    \path 
    (ACT2)
    edge[] node[above] {
      \begin{tabular}{c}
        {\color{\colorevent} $\mathsf{dispatch_2}$} \\
      \end{tabular}
    } 
    (RT2)
    ;

    \path 
    (ACT1)
    edge[] node[below] {
      \begin{tabular}{c}
        {\color{\colorevent} $\mathsf{dispatch_1}$} \\
      \end{tabular}
    } 
    (RT1)
    ;

    \path 
    (RT1)
    edge[] node[above] {
      \begin{tabular}{c}
        {\color{\colorevent} $\mathsf{end_1}$} \\
      \end{tabular}
    } 
    (ET)
    ;
    \path 
    (RT1)
    edge[] node[below] {
      \begin{tabular}{c}
        {\color{\colorevent} $\mathsf{arrival_2}$} \\
      \end{tabular}
    } 
    (RT1WT2)
    ;

    \path 
    (RT2)
    edge[] node[above] {
      \begin{tabular}{c}
        {\color{\colorevent} $\mathsf{end_2}$} \\
      \end{tabular}
    } 
    (ET)
    ;
    \path 
    (RT2)
    edge[] node[above] {
      \begin{tabular}{c}
        {\color{\colorevent} $\mathsf{arrival_1}$} \\
      \end{tabular}
    } 
    (AT1RT2)
    ;

    \path 
    (AT1RT2)
    edge[] node[left] {
      \begin{tabular}{c}
        {\color{\colorevent} $\mathsf{preemption_2}$} \\
      \end{tabular}
    } 
    (RT1WT2)
    ;

    \path 
    (RT1WT2)
    edge[] node[above] {
      \begin{tabular}{c}
        {\color{\colorevent} $\mathsf{end_1}$} \\
      \end{tabular}
    } 
    (ET1WT2)
    ;

    \path 
    (ET1WT2)
    edge[] node[left] {
      \begin{tabular}{c}
        {\color{\colorevent} $\mathsf{dispatch_2}$} \\
      \end{tabular}
    } 
    (RT2)
    ;

    \path 
    (ET)
    edge[] node[left] {
    } 
    (IDLE)
    ;




 
 





  \end{tikzpicture} 
		}
	\label{fig:scheduler}
	}
	
	\caption{The modelling framework for a real-time system}
        \label{fig:model-rts}
\end{figure}
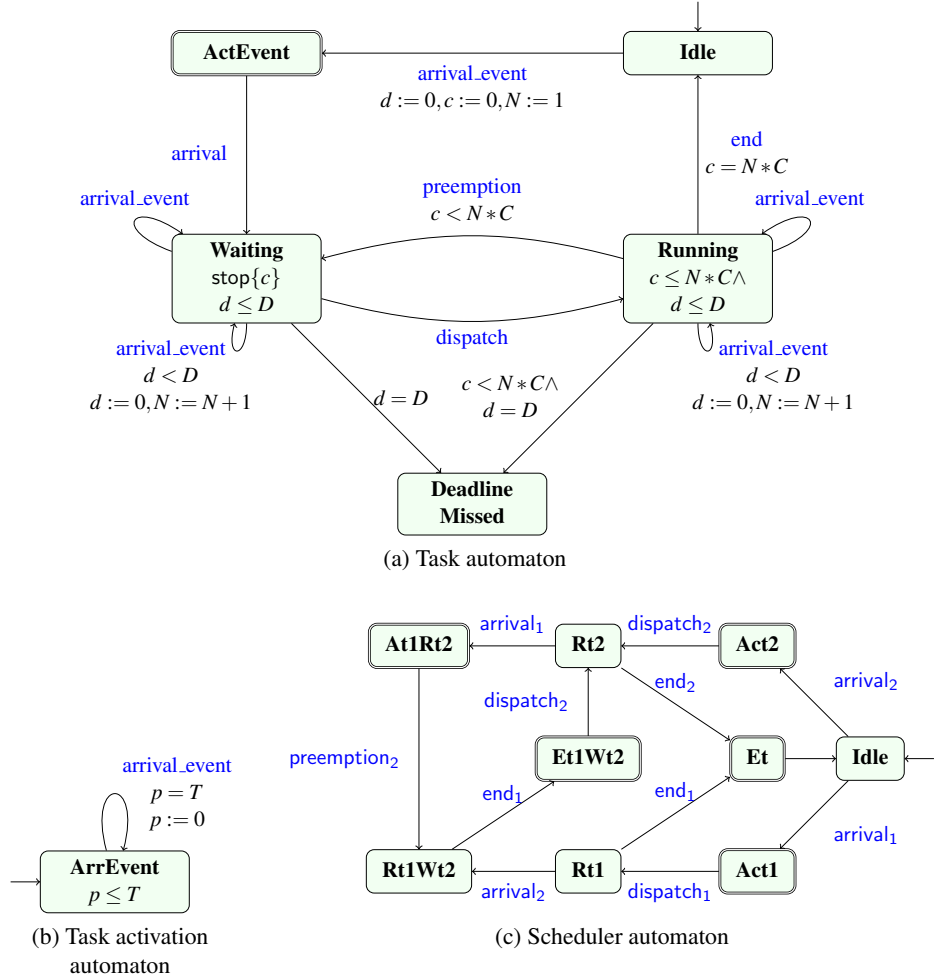

Each task is modelled using a task automaton.
Such a task automaton is shown in Figure~\ref{fig:task}.
Each task automaton contains two (local) continuous clock variables~$c$ and~$d$.
\ea{I changed a bit, and added that clocks are local (which I am almost sure of, right?)}\gl{yes}
Clock~$c$ counts the execution of the task and clock~$d$ counts the time passed since last job arrival.
Since we consider generic activation patterns (periodic, sporadic or arrival curves), a new instance may be activated while the previous ones have not yet completed.
Hence, there could be several active jobs from the same task at the same time.
A discrete\footnote{%
	Discrete variables are not part of the original PTA/PSA formalisms, but can seen as syntax sugar to increase the number of discrete states (locations).
	Such discrete variables are supported by most tools for (parametric) timed automata.
} variable $N$ is used to count the number of simultaneous active instances for the task.

Initially, a task is in location \textsf{Idle}. The synchronisation 
label \textsf{arrival\_event} notifies that a new instance from this 
task is activated and triggers a transition to a committed location 
\textsf{ActEvent}. A committed location is a location where time elapsing
is not allowed, represented graphically using a double circle location.
The label \textsf{arrival} is used between a task 
and the scheduler. The task will then go to location \textsf{Waiting} 
and wait there for the scheduler's decision whether to occupy the 
CPU. If a task has the highest priority among the active tasks in the 
system, the scheduler will send \textsf{dispatch} to trigger the 
transition from \textsf{Waiting} to \textsf{Running}. While a task is in 
\textsf{Running}, the scheduler could revoke the CPU for a 
higher priority task through synchronisation label 
\textsf{preemption}.

Clock $d$ always progresses and the execution time 
clock variable $c$ is stopped if a task is waiting. When a task is 
waiting for the CPU or running on the CPU, to react to new 
activations, it will non-deterministically choose to increase the 
counter~$N$ of active  instances by 1.
When a job misses its 
deadline ($d=D$) before completing its execution, it will go to 
\textsf{DeadlineMissed}.
When a task finishes its execution ($c=N*C$),
it will go back to initial location \textsf{Idle}.

There could be many different activation patterns for a task, such as 
periodic, sporadic or according to arrival curves.
We only require that the activation automaton synchronises with the task automaton 
on label \textsf{arrival\_event}. As a demonstration, 
Figure~\ref{fig:act} shows the activation model for a periodic task. 
Every period $T$, the automaton sends the signal 
\textsf{arrival\_event} to inform the arrival of a new job. 

In this paper, we assume tasks are scheduled according to a fixed-priority 
fully-preemptive scheduler (FPPS). The scheduler automaton synchronises with 
the tasks and decides which task will occupy the CPU at each time.
The structure of the automaton is completely fixed given a number of tasks. 

Figure~\ref{fig:scheduler} shows a scheduler for two tasks, where task $1$ 
has higher priority. The scheduler automaton can be expanded in 
a similar form to deal with a task set with more tasks. In the scheduler 
automaton, the labels \textsf{arrival}, \textsf{dispatch}, 
\textsf{preemption} and \textsf{end} are the same as in task automaton;
we append a label with index $i$, e.g.\ $\mathsf{end_i}$, 
to denote that this label synchronises with task~$i$.
The convention we use for naming the location encodes the status of the tasks: \textsf{Rtx} means the task $\tau_x$ is running; \textsf{Atx} means task $\tau_x$ is just activated; \textsf{Wtx} means task $\tau_x$ is waiting; \textsf{Et} is saying the task just finished its execution.

\section{Parametric Schedulability Analysis of Real-Time Components}
\label{sec:parametric-analysis}




\subsection{Convergence Problem}
\label{sec:convergence}

We first show that the application of the inverse method \IM{} to a system 
with parametric task activations does not yield satisfactory results.
Consider a task set with two periodic tasks $\tau_1=(31,T_1,T_1), 
\tau_2=(49, T_2, T_2)$ with implicit deadlines (i.e.\ deadlines always equal 
to periods). If we use \IM{} with initial values $T_1 = 60$ and $T_2 = 120$, 
respectively, the final constraints
obtained will be $T_1=60$ and $T_2=120$.
That is, the result produced by \IM{} is a single point, the initial 
valuation.

Such result is caused by an important property of the schedule.
The inverse method synthesises a set of constraints that delimit the values 
for the parameters that result in the same exact traces as the initial 
valuation. The schedule generated by a set of periodic real-time tasks
is itself periodic with period $H$ (also called \emph{hyperperiod}).
In particular, the sequence of scheduling events 
repeats itself every $H$, and different $H$ will result in different 
traces of task execution. The hyperperiod can be computed as the 
least common multiple of all task periods: $H = \mathsf{lcm}(T_1, 
\ldots, T_n)$. When periods are parametric, and since function 
$\mathsf{lcm}()$ is highly non linear, a small variation on one period 
can cause very large variations in the hyperperiod.
For example, consider 
the two previous tasks with initial valuation of the periods $T_1 = 
60$ and $T_2 = 120$, respectively. Their hyperperiod is $120$. When 
we increase the second period to $121$, the hyperperiod becomes 
$7260$. Clearly, in this second case the traces are much longer and 
contain many more events. This explains why \IM{} only converges to the 
initial valuation. 

Of course, things become even more complex when considering generic 
arrival patterns. The next section solves this convergence problem 
by exploiting a well-known result from classical scheduling theory.

\subsection{An Improved Model of the System}
\label{sec:idle-model}

As discussed in Section~\ref{sec:convergence}, it is infeasible to 
apply \IM{} directly to a system model with parametric arrival patterns. 
We will try to avoid this situation by adapting the system automaton 
(Figure~\ref{fig:model-rts}) by exploiting the concept of 
\textsf{critical instant}. 

For a set of periodic or sporadic tasks scheduled by FPPS on a single 
processor it is possible to define a \emph{critical scenario}, which is the 
situation that arises when all tasks are simultaneously activated 
(\emph{critical instant}) and every task $\tau_i$ generates subsequent jobs 
as soon as it is allowed. According to the seminal work by Liu and 
Layland~\cite{liu73}, the worst-case response time of a task can be found in 
the \emph{busy period} (i.e.\ interval in which the processor is continously 
busy) that starts at the critical instant. 

This means that, if we want to check the schedulability of a set of 
periodic or sporadic real-time tasks, it is sufficient to activate 
all tasks at time zero and check that no deadline is missed in the 
first busy period starting at time 0. Therefore, as soon as the 
processor becomes idle we can stop our search. 

In the system automaton in Section~\ref{sec:model-rts}, each trace 
corresponds to a possible schedule of the task set. However, we now 
know that to check the schedulability of a task set, it is sufficient to 
analyse traces starting from the critical instant till the first idle
time in CPU. So, we adapt the 
system automaton as follows:
\begin{itemize}
\item The task activation automaton is required to release its first 
job at time 0 and it will emit the subsequent jobs as fast as the 
task is allowed;
\item In the scheduler automaton, after all tasks complete their execution, 
instead of going back to \textsf{Idle}, it will transit from \textsf{Et} to 
a new location \textsf{Stop}, where this is no outgoing edge.
\end{itemize}

The first point is used to simulate the worst-case behaviour of tasks 
at the critical instant. 
Rather than going to \textsf{Idle} and waiting for new task releases,
the scheduler automaton (also the system automaton) simply stops. 
We call this adapted scheduling model as the \emph{idle-time scheduler 
automaton}.

\begin{figure}
\begin{center}
\includegraphics[width=7cm]{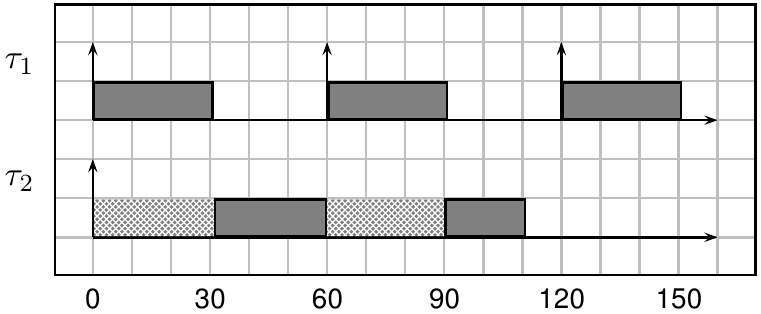}
\end{center}
\caption{Schedule of the first busy period of the example task set}
\label{fig:schedule}
\end{figure}

The idle-time scheduler automaton actually simulates the longest \emph{busy 
period}, which starts from the critical instant and ends at the first idle 
time of the processor. The length of this busy period 
depends both on the execution time and on activation periods of the 
tasks. However, the dependence from the periods is not so strong as 
with the hyperperiod. Let us consider again the previous set of two 
periodic tasks $\tau_1 = (C_1 = 31, T_1 = D_1 = 60)$ $\tau_2 = (49, 120, 
120)$. The schedule for the first busy period is shown in Figure 
\ref{fig:schedule}. Task $\tau_1$ executes twice before the first instance 
of $\tau_2$ can complete. 

The length of the busy period in this case is $2C_1 + C_2 = 111$. By doing 
some simple calculation, it is easy to see that changing $T_1$ to any value 
in $[56,79]$ does not change the sequence of events in the busy period: in 
facts, for any value of $T_1$ in that interval, $\tau_1$ will still execute 
two times before the first instance of $\tau_2$ completes. Also, changing 
$T_2$ to any value $T_2 \geq 112$ does not change the busy period. 

\begin{figure}
\centering
\includegraphics[scale=0.4]{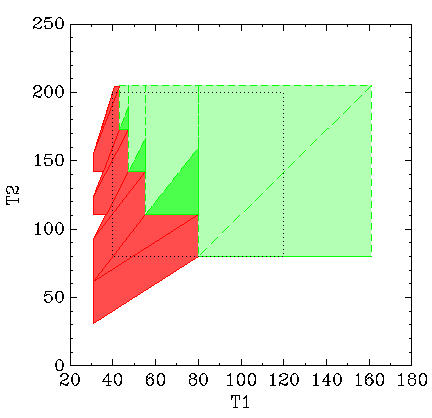}
\caption{Constraints on $T_1$ and $T_2$ obtained by the behavioural cartography}
\label{fig:test1}
\end{figure}

Hence, we can apply \IM{} on the new model and avoid the 
convergence problem as in Section~\ref{sec:convergence}.
Let us assume $T_1\in[40,120]$, $T_2\in[80,200]$ and let us apply the 
behavioural cartography to obtain the constraint space of $T_1, T_2$ that 
keeps the task set schedulable.
The result is given in Figure~\ref{fig:test1} in a graphical form. 
The red part (on the left) is the constraint space on $T_1$ and $T_2$ in 
which the system misses $\tau_2$'s deadlines, whereas the green part (on the 
right) is where no deadline is missed.

When applying the behavioural cartography to the idle-scheduler automaton, 
there may exist a combination of parameters that cause the system to 
go into overload, i.e.\ there will be no idle time in the schedule. 
For example, in case of periodic tasks, this happens when the total 
system utilisation is such that $\sum_{i=1}^n \frac{C_i}{T_i} > 1$.
In the 
previous example, $(T_1 = 40, T_2 = 80)$ is one such point. Of 
course, this will surely cause a deadline miss, because it means that 
the total amount of work to be performed (utilisation) exceeds the 
amount of available processor time. 

To solve this case, we put an upper bound on the maximal depth of the 
traces computed by \IM{}. This bound is always computable in the 
case of periodic real-time tasks, and corresponds to computing an 
upper bound to the time where a deadline miss will happen. A method 
for computing such a bound can be built by using the concept of 
\emph{demand bound function}~\cite{BHR90}.

\subsection{Applicability of the Idle-Time Scheduler}
\label{sec:validity}

It is possible to prove that the concepts of critical instant and 
maximal busy periods are valid also when considering tasks activated 
by generic arrival curves \cite{thiele2000real}. In particular, the 
critical scenario corresponds to the time instant in which all tasks 
are activated with their initial burstiness (critical instant), and 
their successive instances arrive as soon as possible without 
violating their arrival curves. Then, the worst-case response time 
can be found in the busy period starting at the critical instant and 
corresponding to the critical scenario. Therefore, we will use the 
same technique also for generic arrival curves.

\begin{figure}
  \centering
   
  \begin{tikzpicture}[->, =stealth!]

    \node[initial left, initial text={n:=0}, state, text width=2.5cm, 
anchor=center, accepting] 
(BURSTING) at (0,0) {      
      \begin{tabular}{c}
        \textbf{Bursting} \\
      \end{tabular}
         
    };
    \node[state, text width=2.5cm, 
anchor=center] 
(ARREVENT) at (5,0) {      
      \begin{tabular}{c}
        \textbf{ArrEvent} \\
        $p\leq P$
      \end{tabular}
         
    };

    \path
    (BURSTING)
    edge [above, out=105, in=75, looseness=12] node[left=2mm] {
      \begin{tabular}{@{} c @{}}
         {\color{\colorevent} arrival\_event}\\
         $ n < N^{u} - 1$ \\
         $ n := n+1 $ 
      \end{tabular}
    }
    (BURSTING)
    ;
    \path
    (BURSTING)
    edge [] node[above] {
      \begin{tabular}{@{} c @{}}
         {\color{\colorevent} arrival\_event}\\
         $ n = N^{u} - 1$ \\
         $ p := 0 $ 
      \end{tabular}
    }
    (ARREVENT)
    ;
    \path
    (ARREVENT)
    edge [above, out=105, in=75, looseness=12] node[right] {
      \begin{tabular}{@{} c @{}}
         {\color{\colorevent} arrival\_event}\\
         $ p = P $ \\
         $ p := 0 $ 
      \end{tabular}
    }
    (ARREVENT)
    ;

  \end{tikzpicture}
  \caption{Arrival curve automaton}
  \label{fig:curve}
\end{figure}
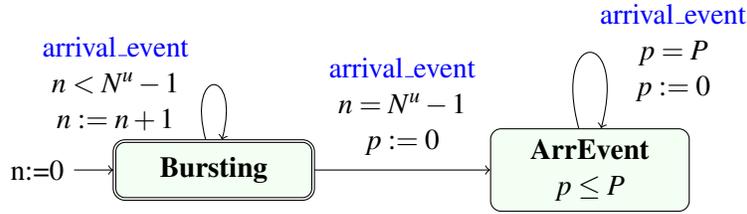

In Figure \ref{fig:curve} we show the simple PSA model for a parametric 
periodic arrival curve described by Equation~\ref{eq:param-arr-curve}. 
Initially, the arrival curve automaton is in a committed location 
\textsf{Bursting} with $n=0$, where $n$ is a discrete variable counting the 
number of initial client requests.
The automaton emits $N^{u}$
activations for a task ($\tau_2$ in our case) within 0 time elapse and  then 
moves to location $\textsf{ArrEvent}$ where is starts behaving as a periodic 
activation automaton as in Figure~\ref{fig:act}, and produces activations 
events every $P$. 

For other different task models there is no critical instant. For example, 
when considering periodic tasks with initial offset different from zero, 
there is no worst-case scenario in the schedule. Instead, it is necessary to 
analyse all busy periods in the interval $[0, 2H + 
\Phi_{\mathsf{max}}]$, where $\Phi_{\mathsf{max}}$ is the largest initial 
offset~\cite{LW82}. 

Given a task set $\mathcal{T}$ of periodic real-time task with offsets, 
we can build a task set $\mathcal{T}^\prime$ that contains the same 
tasks with the same parameters except that their initial offsets are all 
set  to zero. In this case, it is possible to prove that, if 
$\mathcal{T}^\prime$ is schedulable, then also $\mathcal{T}$ is 
schedulable. However, the converse does not hold. Therefore, it is 
possible to perform a parametric analysis of $\mathcal{T}^\prime$ using 
our idle-time scheduler, and the set of values of the parameters produced 
by the analysis is a subset of the set of valid parameters for the 
original system $\mathcal{T}$. A more precise analysis requires point-by-
point exploration of the parameter space.

Finally, in this paper we assume that task are independent from each 
other, and do not self-suspend waiting for other events different from 
the activation event. An example of self-suspending task is a task that 
performs a remote procedure call, and self-suspends waiting for the 
response. Again, in this case there is not a single critical scenario for 
the task set, therefore our simplified model cannot be used. 

\section{Towards Timed Interfaces}
\label{sec:interface}

In this section we show how it is possible to define a \emph{timed 
interface} of a real-time component using parametric analysis.

Consider the system of Figure~\ref{fig:component-representation}: it 
consists of 3 tasks $\tau_1, \tau_2$ and $\tau_3$
running on a single processor with FPPS. A task with smaller index has 
higher priority. $\tau_1$ and $\tau_3$ are periodic tasks with $\tau_1 = 
(C_1 = 2, D_1 = 8, T_1 = 8)$ and $\tau_3 = (C_3 = 20, D_3 = 50, T_3 = 
50)$. Task $\tau_2$ has $C_2=5$ and implements the method provided in the 
interface. We assume that this component is linked to a local networks, 
and task $\tau_2$ receives the requests from clients running on other 
nodes of the network. We would like to know how many clients can ask 
requests to the system, with which frequency, and the maximum delay that 
is going to pass from the request to the response. Therefore, we need to 
study the possible activation patterns of task $\tau_2$ and its worst-case 
response time.
For modelling the activation patterns, we use a 
parametric arrival curse as described by Equation~\ref{eq:param-arr-curve}.
For example, $N^{u} = 2$ and $P=100$ means that we can connect 
at most 2 independent clients, and that between any two consecutive 
requests after the first two there must be at most $100$ units of time. 

Both $N^{u}$ and $P$ are parameters we are going to synthesise with our 
parametric analysis. Another parameter is the delay (deadline) $D_2$ of 
$\tau_2$. We are interested in the parameter space that guarantees all 
the tasks are schedulable.

First, we construct the activation automaton for $\alpha(t)$ as in
Figure~\ref{fig:curve}. Following the method described in 
Section~\ref{sec:model-rts}, and using the idle-time scheduler automaton, 
we then compose the final automaton.

Given that $C_2=5$, it is easy to see that the burst ($N^{u}$) of the 
arrival curve automaton cannot be larger than $3$, otherwise $\tau_3$ 
will be doomed to miss its deadline, because $D_3 < C_3 + 4C_2 + 5C_1$. 
Additionally, we assume $P$ and $D_2$ lie in following intervals:
\begin{equation*}
P\in[20,50], \,\,\, D_2\in[10,50]
\end{equation*}
$N^u$ is a discrete parameter that must be treated separately from the 
other parameter. Our strategy is to instantiate $N^{u}$ with 1, 2 and 3 
individually and apply \IM{} to each case in order to synthesise 
constraints over $P$ and $D_2$ that keep the system schedulable. The 
resulted parameter spaces for the three cases are visualised in 
Figure~\ref{fig:curve-results}. 

We can use these values to build a \emph{timed interface 
specification} for the component. 
\begin{itemize}
\item the number of distinct independent clients that can be connected to the service must respect the constraint $1 \leq N^u \leq 3$;
\item Depending on the number of clients, the relationship between minimum request period $P$ and worst case response time $D_2$  is specified in Table \ref{tab:interface}.
\end{itemize}

\subsubsection*{Reducing the number of regions}

\begin{figure}
	\centering
	\subfloat[$N^{u}=1$]{
	\includegraphics[width=0.32\textwidth]{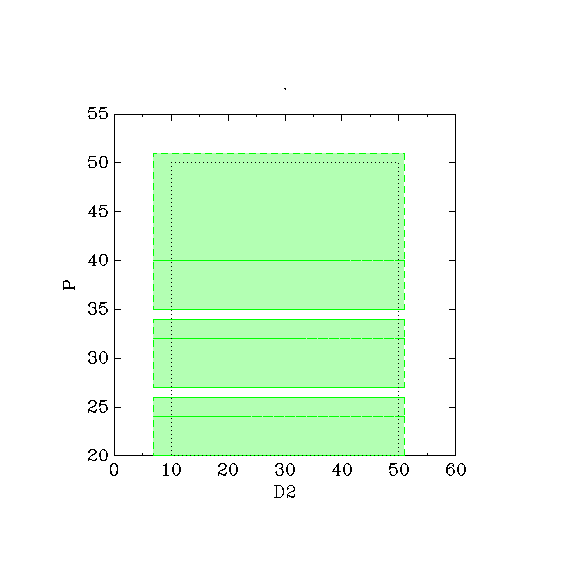}
	}
	\hfill
	\subfloat[$N^{u}=2$]{
	\includegraphics[width=0.32\textwidth]{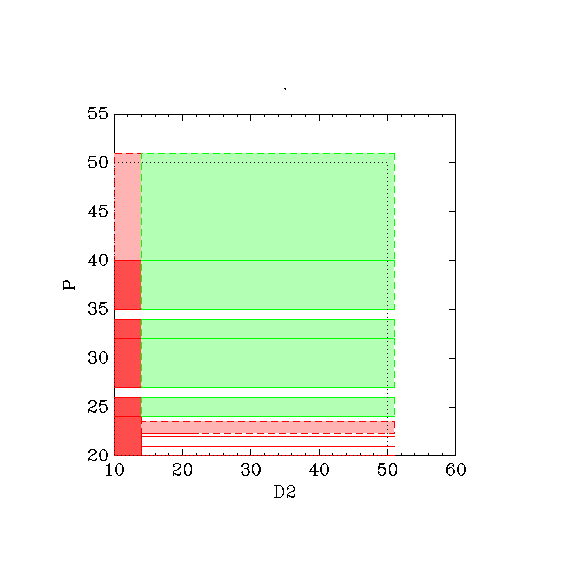}
	}
	\hfill
	\subfloat[$N^{u}=3$]{
	\includegraphics[width=0.32\textwidth]{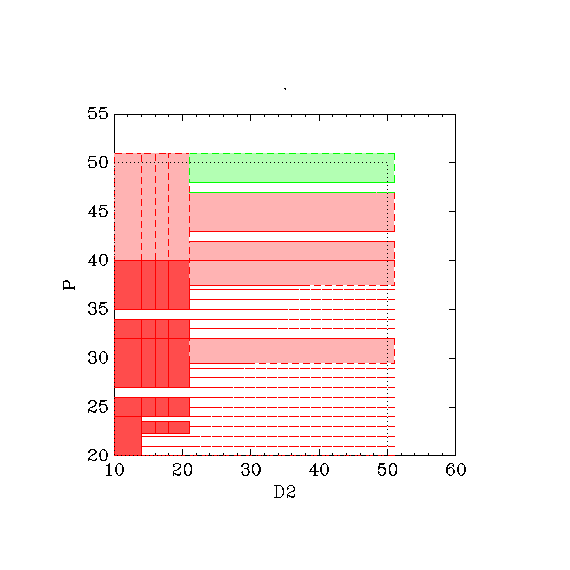}
	}
	\caption{Parameter space (green) for $N^{u}$, $P$ and $D_2$}
	\label{fig:curve-results}
\end{figure}

As it is possible to see in Figure \ref{fig:curve-results} and in Table 
\ref{tab:interface}, the parameter space returned by \IM{} consists of a 
set of disjoint tiles. Each tile is a convex region and the resulting 
interface is the union of (maybe a large number of) these convex regions. 
Such an interface may not be easy to use due to the large number of disjoint regions. 

In some cases, it is possible to perform a ``merge'' operation between 
the tiles, as explained in~\cite{AFS13atva}, in order to reduce the 
number of convex regions composing the final interface. Two convex 
regions are mergeable if their convex hull equals to their union. Given 
tiles returned from \IM{}, we repeatedly replace mergeable tiles, 
satisfying this condition, with their union till there are no mergeable 
tiles. 
If we restrict ourselves to integers solutions, we may further merge adjacent tiles. For example, the constraint $(20 \leq P \leq 26)$ can be 
merged with $(27 \leq P \leq 34)$, thus obtaining $(20 \leq P \leq 34)$. 
We are currently investigating efficient methods for automatically merging tiles resulting from \IM{} cartography.
\ea{This should actually be easy to do; remind me to talk about it next time we meet!}

\begin{table}
	\centering
	\begin{tabular}{|c|c|}
	\hline
	$N^{u}=1$     &   when $(20\leq P\leq 26)\vee  (27\leq P\leq 34) \vee (35\leq P\leq 50) \rightarrow D_2^{min} = 10$ \\\hline
	$N^{u}=2$     &   when $(24\leq P\leq 26)\vee  (27\leq P\leq 34) \vee (35\leq P\leq 50) \rightarrow D_2^{min} = 14$ \\\hline
	$N^{u}=3$     &   when $(P=47) \vee (48\leq P\leq 50) \rightarrow D_2^{min} = 21$ \\\hline
	\end{tabular}
	\caption{The final interface}
	\label{tab:interface}
\end{table}

%
%
%
%
%

\section{Conclusion and Future Work}
\label{sec:conclusions}

In this paper we have presented a PTA model of a real-time systems scheduled 
by FPPS. We have shown how to perform a parametric analysis using \IM{} with 
a specific model of the scheduler that stops at the first idle time. 
Finally, we have shown how to use parametric analysis for the design and the 
specification of the interface of a real-time component. 

We wish to continue along this line of research and investigate about the 
possibility to systematically use parametric analysis for interface 
specification. We are currently investigating efficient methods for reducing 
the complexity of the set of regions produced by \IM{}, either by using more 
sophisticated merging techniques, or by using conservative approximations. 
Also, we plan to extend the analysis to more complex task models like self-suspending tasks and task dependencies.

More specifically on the parameter synthesis techniques, it would be 
interesting to reuse some technique for integer parameter synthesis 
recently proposed in~\cite{JLR13};
on the negative side, only integer points are synthesised, thus preventing 
the interpretation of the result for robustness analysis (in the sense of 
infinitesimal variations of the parameters);
on the positive side, these techniques are efficient and guaranteed to 
terminate.
Also, combining the inverse method with IC3~\cite{CGMT13} is an interesting 
future direction of research.

A more general (and challenging) objective is also to be able to derive (possibly non-linear) constraints relating the discrete and continuous parameters, e.g.\ relating the number of clients (``$N^u$'' in Section~\ref{sec:interface}) with the timing parameters (``$P$'' and ``$D_2$'' in Section~\ref{sec:interface}).

\subsection*{Acknowledgment}
We would like to thank anonymous reviewers for their useful comments.

\ea{Note for myself for the next version of the paper: say something about changing IM to avoid the problem in 5.1, or use a different algorithm.}

\bibliographystyle{eptcs}
\bibliography{biblio}

\end{document}